\documentclass[aps,prl,twocolumn,superscriptaddress,showpacs,pagewise]{revtex4-2}
\usepackage{amsmath}
\usepackage[dvips,xdvi]{graphicx}
\usepackage{amssymb}
\usepackage{epsfig}
\makeatletter
\providecommand{\@LN}[2]{}
\makeatother

\begin{document}
\title{Quantum critical dynamics in a spinor Hubbard model quantum simulator}
\author{J. O. Austin}
\author{Z. Chen}
\author{Z. N. Shaw}
\affiliation{Department of Physics, Oklahoma State University, Stillwater, Oklahoma 74078, USA}
\author{K. W. Mahmud}
\email{Present address: Quidient LLC, Columbia, MD 21046}
\affiliation{Joint Quantum
Institute, University of Maryland, College Park, Maryland 20742, USA}
\author{Y. Liu}
\email{Electronic address: yingmei.liu@okstate.edu} \affiliation{Department of Physics, Oklahoma State University,
Stillwater, Oklahoma 74078, USA}
\date{\today}

\maketitle

\textbf{Physical modeling of three-dimensional (3D) strongly correlated many-body systems, especially their critical
dynamics across quantum phase transitions, is a fast-moving research frontier with immediate applications spanning from
adiabatic quantum state preparations to developing novel materials~\cite{StamperKurnRMP,blochRMP,dynamicsRMP,
Chen,Braun,JiangLattice,ChenQuench,Dziarmaga,LatticeSpinor}. Impeded by the intrinsic complexity of these systems and
limitations of existing numerical techniques, theoretical studies have been performed mainly in homogeneous systems and
lower dimensions~\cite{Braun,Shimizu,Gutzwiller07,Gutzwiller15,1Dtheory1,1Dtheory2,1Dtheory3,zurek2017,blochRMP,dynamicsRMP}.
Here we experimentally demonstrate that such complex dynamics can be efficiently studied in 3D spinor Bose-Hubbard model
quantum simulators, consisting of lattice-confined spinor condensates. Possessing spin degrees of freedom and exhibiting
magnetic order and superfluidity, these 3D quantum simulators are highly
programmable~\cite{StamperKurnRMP,blochRMP,Ho1998,Adilet2003,spinorquench,JiangLattice,ZhaoSinglet,ChenQuench}. We find
novel dynamics and scaling effects beyond the scope of existing theories at superfluid-insulator quantum phase
transitions, and highlight spin populations as a new observable to probe these dynamics. We also conduct numerical
simulations in lower dimensions using time-dependent Gutzwiller approximations, which qualitatively describe our
observations.}

The 3D spinor quantum simulators realized in this work, directly implementing the important Bose-Hubbard (BH) model, are
well-isolated quantum systems consisting of antiferromagnetic spinor Bose-Einstein condensates (BECs) in cubic optical
lattices~\cite{spinorquench}. When lattice-trapped bosons are suddenly quenched from the superfluid (SF) phase to the
Mott-insulator (MI) phase, the BH model is nearly integrable, yielding non-thermal steady states preserving memory of the
initial state~\cite{Dziarmaga,1Dtheory3}. One distinctive feature of these 3D quantum simulators is their ability to tune
the nature of SF-MI quantum phase transitions, i.e., spinor gases can cross first-order (second-order) SF-MI transitions
when the quadratic Zeeman energy $q$ is smaller (larger) than the spin-dependent interaction energy $U_2$ in even Mott
lobes~\cite{StamperKurnRMP,Adilet2003,spinorquench,Gutzwiller15,ZhaoSinglet,JiangLattice}. $U_2$ also enables spin-mixing
oscillations among multiple spin components and generates ferromagnetic and antiferromagnetic order along with various
magnetic states including spin singlets and nematic
states~\cite{StamperKurnRMP,Ho1998,Adilet2003,spinorquench,ZhaoSinglet}. The 3D quantum simulators have thus been
suggested as an ideal platform for studying quantum coherence, long-range order, magnetism, quantum phase transitions, and
symmetry breaking~\cite{StamperKurnRMP,spinorquench,JiangLattice,ChenQuench,ZhaoSinglet}.

In this paper, we experimentally explore these quantum critical dynamics by demonstrating asymmetric impulse-adiabatic
regimes~\cite{zurek2017,Schutzhold} as spinor gases are quenched across superfluid to spinor Mott-insulator quantum phase
transitions at different speeds. Dynamics of phase transitions have been a rich but challenging research theme. One
paradigm of dynamical studies is the Kibble-Zurek mechanism (KZM), which characterizes dynamics into two regimes: the
impulse regime, where a system is frozen due to diminished energy gaps and slowed equilibrations in the vicinity of phase
transition points, and the adiabatic regime, where a system with considerable energy gaps progresses
adiabatically~\cite{KZM,KZM4,KZM2,KZM3,Dziarmaga}. Our observations, going beyond the adiabatic-impulse-adiabatic regimes
studied in many publications~\cite{KZM,KZM2,KZM4,KZM3,zurek2017,Dziarmaga,beyondKZ20,Chandran}, confirm the impulse regime
is enlarged to cover the entire SF phase and part of the MI phase due to negligible excitations within the SF
phase~\cite{zurek2017,Schutzhold}. We also find novel quench dynamics and scaling effects at SF-MI quantum phase
transitions, which are outside the framework of existing theories including KZM. The observed power law scaling exponents
are independent of the nature of SF-MI transitions. In addition, our data present advantages of using spin populations as
a new observable for probing nonequilibrium dynamics, as they are less sensitive to lattice-induced heating than
observables (e.g., visibility) used in scalar bosons. While numerical simulations in 3D multi-component systems are
prohibitively difficult to perform for further verification of our results, we conduct theoretical simulations in lower
dimensions and find qualitative agreements with the experiment.

\begin{figure*}[tb]
\includegraphics[width=166mm]{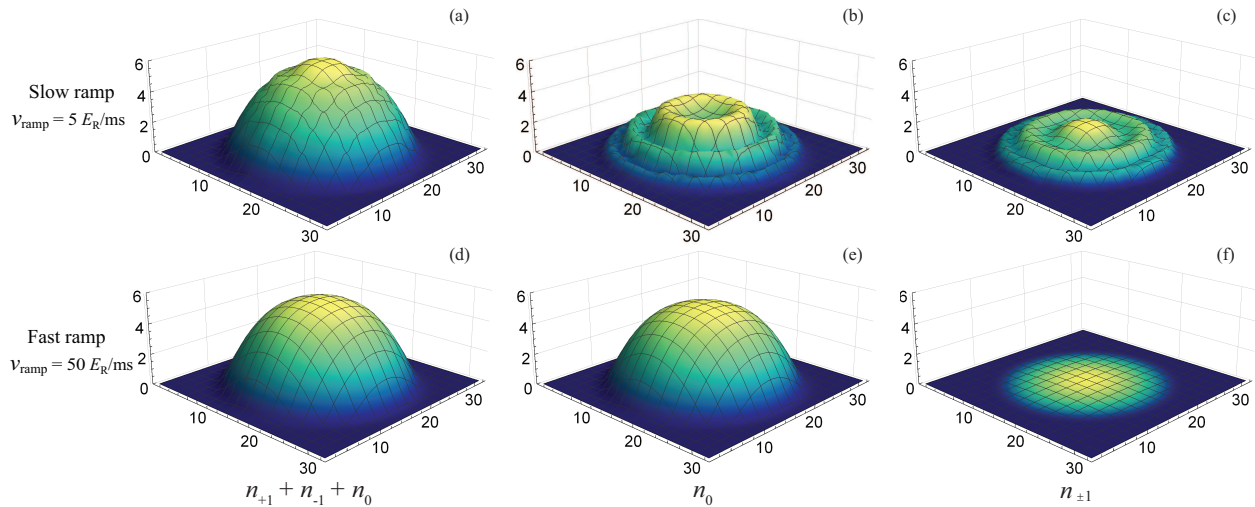}
\caption{(a,b,c) Predicted density profiles of all atoms, the $m_F=0$ atoms, and the $m_F=\pm1$ atoms after a slow quench
from a shallow lattice of $u_L=2E_R$ to a deep lattice at 40$E_R$, respectively. These are derived from our numerical
simulations for 2D lattice-trapped sodium spinor gases of $n_{\rm peak}=6$ at $q/h=42$~Hz using the BH model and the
Gutzwiller approximation (see Methods). Here $n_{\rm peak}$ is the peak occupation number per lattice site, $q$ is the
quadratic Zeeman energy, and $h$ is the Planck constant. (d,e,f) Similar to Panels (a,b,c) but after a fast lattice
quench.} \label{theory}
\end{figure*}

We start each experimental cycle with a $F=$~1 antiferromagnetic sodium spinor BEC at its SF ground state, the
longitudinal polar state, in which $\rho_0=1$ and $m=0$. Here $\rho_{m_F}$ is the fractional population of atoms in the
spin-$m_F$ state and $m=\rho_{-1}+\rho_{+1}$ is the magnetization. We then linearly quench the depth $u_L$ of cubic
lattices, which results in an exponential increase in the ratio of the spin-independent interaction $U_0$ to the hopping
energy $J$. Similar to scalar bosons, spinor gases can cross SF-MI transitions when $U_0/J$ becomes larger than critical
values~\cite{JiangLattice,ChenQuench,spinorquench,Adilet2003,spinorquench,ZhaoSinglet,Gutzwiller15}. The static and
dynamic properties of the spin-1 BH Hamiltonian are computed using the Gutzwiller approximation (see Methods).

\begin{figure*}[tb]
\includegraphics[width=160mm]{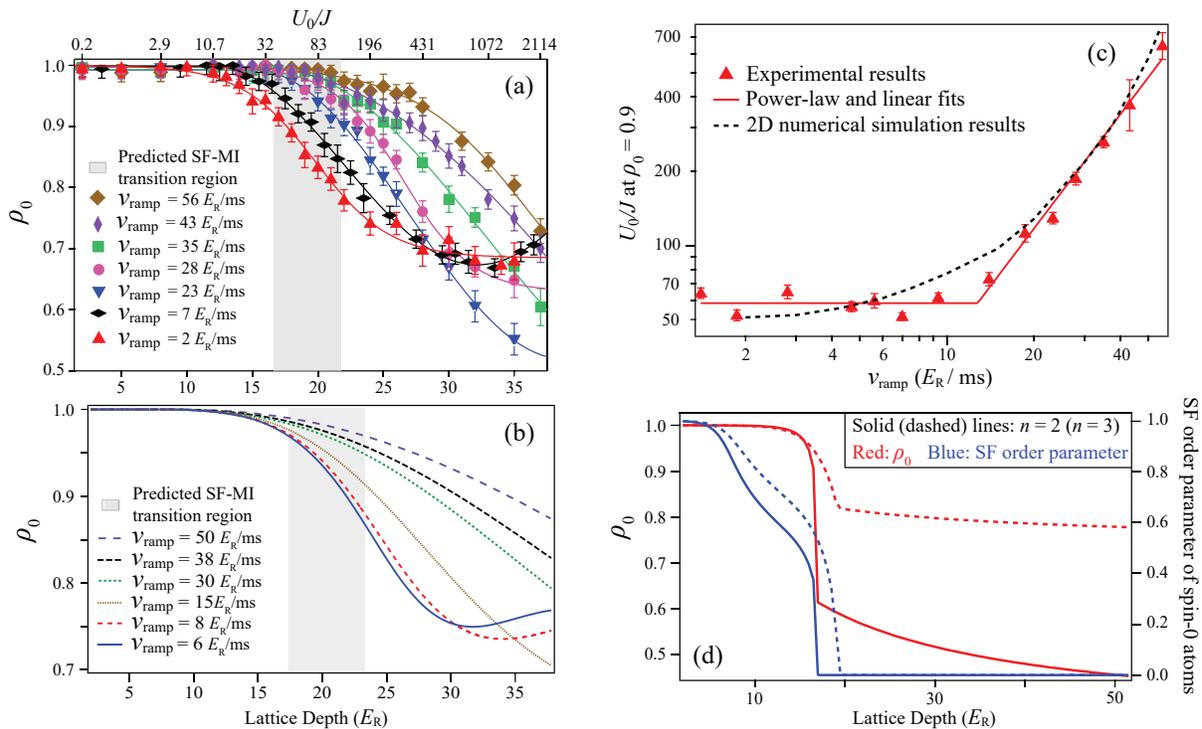}
\caption{(a) Observed spin dynamics at various lattice quench speeds $v_{\rm ramp}$ in magnetic fields of $q/h=42$~Hz. The
$7E_R/\mathrm{ms}$ curve (in black) is offset horizontally by $1.5E_R$ for visual clarity. Solid lines are fitting curves
to guide the eye, and the shaded gray area represents the predicted SF-MI transition region for the filling factor of
$1\leq n\leq 6$. The top horizontal axis lists the corresponding $U_0/J$. (b) Similar to Panel (a) but derived from the BH
model for 2D lattice-trapped sodium spinor gases (see Methods). (c) Markers correspond to $U_0/J$ extracted from Panel~(a)
at the cutoff of $\rho_0=0.9$ as a function of $v_{\rm ramp}$. The solid line represents a power-law fit of the data in
fast quenches and a linear fit in slow quenches. The dashed line is the 2D numerical simulation result. (d) Red and blue
lines represent the predicted behavior of $\rho_0$ and the SF order parameter of the spin-0 component after adiabatic
ramps in 3D lattice-trapped sodium spinor gases at $q/h=42$~Hz, respectively. The SF order parameter is nonzero (zero) in
the SF (MI) phase (see Methods and Supplementary Information). Solid (dashed) lines represent the $n=2~(n=3)$ Mott lobe. }
\label{rho0}
\end{figure*}

Figure~\ref{theory} compares density profiles of various spin components, derived from numerical simulations in systems
similar to the experimental system with fewer dimensions, after the lattice depth is quenched from 2$E_R$ to 40$E_R$ at
different speeds. Here $E_R$ is the recoil energy~\cite{spinorquench,JiangLattice}. For slow quenches,
Fig.~\ref{theory}(a)-(c) clearly display the formation of Mott insulating shells and spin structures. Whereas for fast
quenches, Fig.~\ref{theory}(d)-(f) show persistence of SF type density profiles. Lattice gases studied in this paper are
inhomogeneous systems confined by an overall harmonic trap. In these systems, the BH model predicts that SF and MI regions
coexist and many Mott lobes are arranged in a wedding-cake structure (see Fig.~\ref{theory} and Supplementary
Information). Such inhomogeneous systems have been suggested as good candidates for adiabatic quantum state preparations,
because inhomogeneous phase transitions suppress excitations during quenches~\cite{Dziarmaga}.

Our experimental data agree with these predictions, as shown by spin dynamic curves in Fig.~\ref{rho0}(a). During fast
quenches, the observed spin-0 population $\rho_0$ implies that the initial SF state is frozen as the lattice is
quenched across predicted
SF-MI transition points (shaded gray area in Fig.~\ref{rho0}(a)) and then starts to evolve inside isolated lattice sites
in deep lattices. Whereas in sufficiently slow quenches, our data indicate that atoms stay in the instantaneous ground
state during the dynamics and enter the MI phase at a critical $U_0/J$. For intermediate quench speeds, for example the
$7E_R/\mathrm{ms}$ curve in Fig.~\ref{rho0}(a), partial revivals in the spin-0 population may be observed after $\rho_0$
reaches its minimum value. Such revivals are not found at slow and fast quench speeds (see Fig.~\ref{rho0}(a)).
Qualitatively similar results are obtained from our 2D numerical simulations (see Fig.~\ref{rho0}(b)), although it is
harder to realize adiabatic phase transitions in lower dimensional systems where more low-energy states are accessible for
excitations~\cite{Dziarmaga}. Theoretical calculations also predict the frequency of the periodic revivals is determined
by $U_2$ and $q$ for a fixed atom number $n$, while the amplitude of the revivals depends on the quench speed $v_{\rm
ramp}$. The physical limitations of our system, however, prevent us from safely observing full periods of the revivals at
intermediate $v_{\rm ramp}$, as well as the minimum $\rho_0$ in fast quenches ($v_{\rm ramp}\gtrsim28E_R/\mathrm{ms}$),
due to extremely high power requirements on the lattice beams.

We apply a single $\rho_0$ cut-off value to mark the edge of the impulse regime. For example, Fig.~\ref{rho0}(c) is
extracted from the observed dynamics by setting the end of the impulse regime at the lattice depth where $\rho_0=0.9$.
When $U_0/J$ at this lattice depth is plotted against $v_{\rm ramp}$, two distinct dynamical regions emerge (see
Fig.~\ref{rho0}(c)). For relatively slow quenches $(v_{\rm ramp}\lesssim13E_R/\mathrm{ms})$, the impulse regime does not
strongly depend on quench speed. For faster quenches, the length of the impulse regime obeys a power law relationship with
$v_{\rm ramp}$ as expected from a quantum version of the KZM, although the extracted scaling exponent $1.55(8)$ is much
larger than the 0.67 predicted by a simple KZM for lattice-trapped 3D scalar bosons~\cite{Dziarmaga}. The extracted
scaling exponents show no strong dependence on the $\rho_0$ cut-off value as long as the cutoff is sufficiently larger
than the minimum observed $\rho_0$ values. We find a qualitative agreement between the 2D simulation (dashed line in
Fig.~\ref{rho0}(c)) and our data, although the scaling exponents are hard to match between theory and experiment due to
the presence of inhomogeneous systems and difference in dimensionality. The quenches studied in this paper are exponential
in $U_0/J$ whereas the KZM usually applies to quenches linear in system parameters, so our experiments may have revealed
non-universal scaling relations. Our numerical simulations have taken this feature into account, while omitting any
finite-temperature or heating effects.

\begin{figure*}[t]
\includegraphics[width=175mm]{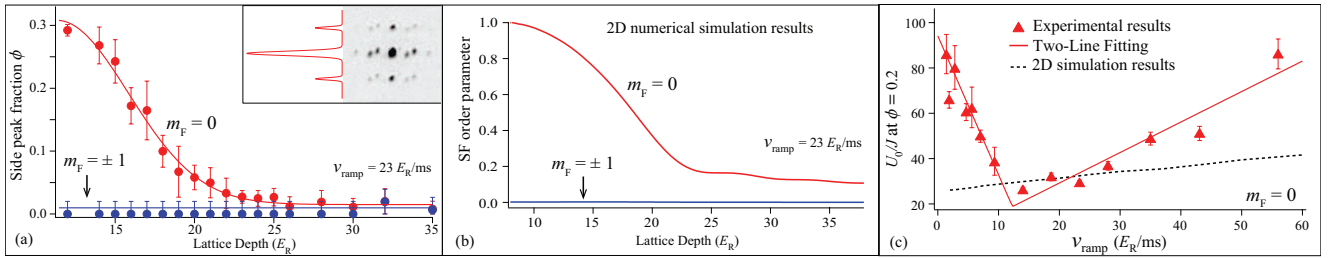}
\caption{(a) Red (blue) markers represent the observed side peak fractions $\phi$ of the $m_F=0$ ($m_F=\pm 1$) spin
components at $q/h=42$~Hz during a lattice quench at an intermediate speed. The red (blue) line is a Gaussian (linear)
fit. The inset shows how we extract $\phi$ from a TOF image (right). The left inset is a density profile, which is created
by integrating along a vertical slice of the TOF image and is then fit with four Gaussian curves. From these fittings we
can extract $\phi$ by taking the area of the two side peaks and dividing by the total area. (b) Similar to Panel (a) but
derived from the BH model for 2D lattice-trapped sodium spinor gases (see Methods). (c) Markers represent $U_0/J$, where
the measured $\phi$ of the spin-0 atoms equals 0.2, extracted from curves similar to the one shown in Panel(a) but taken
at various $v_{\rm ramp}$. The solid line is a two-line fit and the dashed line represents the 2D numerical simulation results.}
\label{order}
\end{figure*}

Both $\rho_0$ and SF order parameters are theoretically good observables for SF-MI phase transitions, due to the
similarities in the topology and boundaries of the phase diagrams derived from SF order parameters and $\rho_0$ when
$n>1$. A typical example at an even $n$ and an odd $n$ is shown in Fig.~\ref{rho0}(d). To emphasize a key advantage of
using spin populations as a new observable to probe SF-MI transitions, we further analyze the dynamics by examining the SF
order parameter, which is proportional to the side peak fraction $\phi$ measured in experiments. $\phi$ is a measure of
coherence of the wavefunction and is proportional to the number of atoms at zero momentum in the spin-dependent momentum
distributions (see Methods)~\cite{JiangLattice,Ketterle2006}. The inset of Fig.~\ref{order}(a) illustrates how we extract
$\phi$. Typical evolutions of $\phi$ for the three spin components during a quench at an intermediate speed are presented
in Fig.~\ref{order}(a). For the spin-0 atoms, $\phi$ increases steadily until it obtains its maximum around $12E_R$ and
then decreases as the lattice gets deeper and the interference pattern disappears. In contrast, the observed $\phi$ for
the spin-$\pm 1$ components always stays at zero throughout the evolution, which indicates $m_F=\pm1$ atoms are only
formed in the MI phase. These observations agree with our theoretical simulations shown in Fig.~\ref{order}(b).

\begin{figure}[t]
\includegraphics[width=82mm]{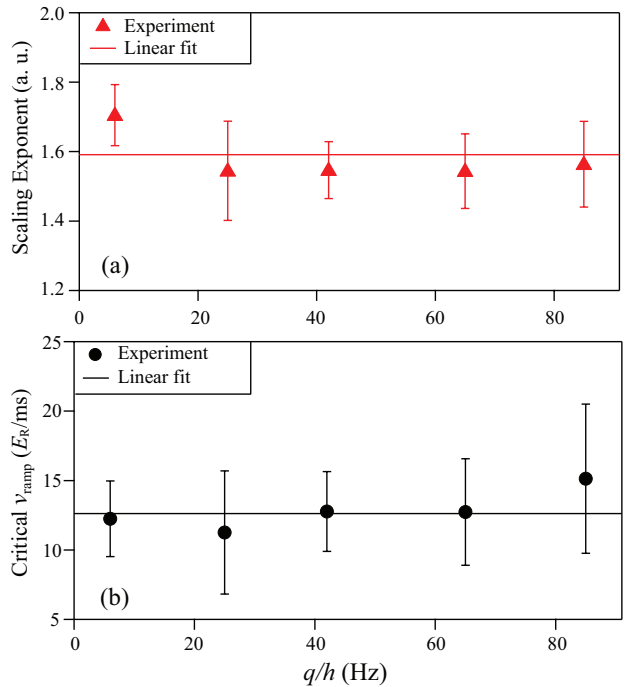}
\caption{(a) The scaling exponent extracted from the power-law fits as illustrated in Fig.~\ref{rho0}(c) at various
quadratic Zeeman energy $q$. (b) The critical lattice quench speed at the intersection of the linear and power-law fits
shown in Fig.~\ref{rho0}(c) as a function of $q$. Solid lines are linear fits.} \label{scaling}
\end{figure}
By defining a cut-off $\phi$ at which the fitting curve for $\phi$ of the spin-0 atoms equals 0.2, two different regions
again emerge, as shown in Fig.~\ref{order}(c). These regions are divided at roughly the same critical $v_{\rm ramp}$
($\approx13E_R/\mathrm{ms}$) as the two regions found in Fig.~\ref{rho0}(c). For fast (slow) quenches where $v_{\rm ramp}$
is faster (slower) than 13$E_R/\mathrm{ms}$, Fig.~\ref{order}(c) shows that the extracted $U_0/J$ linearly increases
(decreases) with $v_{\rm ramp}$, instead of the power law dependence found in Fig.~\ref{rho0}(c). This apparent
discrepancy together with large disagreements between the numerical simulations and the data (see Fig.~\ref{order}(c)) may
confirm SF order parameters are poor observables for non-adiabatic quenches across SF-MI transitions. One possible
explanation is the dynamic heating rate effects upon SF order parameters found in rapid lattice
quenches~\cite{Tiesinga2005,Bloch2006}. This, coupled with the coexistence of SF and MI regions in inhomogeneous systems,
makes it difficult to compare the impulse regimes derived from SF order parameters across varying quench speeds.

To test how different local magnetic fields might affect these results, we repeat these experiments at multiple
quadratic Zeeman energy $q$ within $0<q/h<100$~Hz. We find no significant $q$ dependence in the scaling exponent or the critical
$v_{\rm ramp}$, as shown in Fig.~\ref{scaling}. Because $q\ll U_0$ for our system, the lack of a significant $q$
dependence is consistent with our theoretical understanding of these transitions. From these fittings, we find the scaling
exponent of our system to be $1.6(1)$ and the critical quench speed at which the impulse regime begins to significantly
vary with quench speed to be $13(2)~E_R/\mathrm{ms}$. Because SF-MI phase transitions are first order in even Mott lobes
at $q< U_2\approx h\times 50$~Hz while second order in odd Mott lobes at any
$q$~\cite{StamperKurnRMP,Adilet2003,spinorquench,Gutzwiller15,ZhaoSinglet,JiangLattice}, Fig.~\ref{scaling} implies the
scaling exponents are independent of the nature of the SF-MI phase transitions.

In conclusion, we have experimentally demonstrated novel quench dynamics and scaling effects outside the scope of existing
theories in 3D spinor Bose-Hubbard model quantum simulators. Our data indicate the scaling exponents are independent of
the nature of SF-MI phase transitions. We have also presented the advantages of using spin populations as a new observable
to probe dynamics of SF-MI transitions. While it is prohibitively difficult to compute dynamics of multi-component bosons
in 3D lattices, we have performed numerical simulations in lower dimensions and found qualitative agreements with the
experiment. This work may indicate the 3D quantum simulators could be more powerful than their classical counterparts,
especially in studying many-body quantum dynamics and the interplay of magnetism and superfluidity, although their
effectiveness and accuracy is worthy of further study.
\\

\noindent \textbf{Methods}\\
Each experimental sequence begins by creating a spin-1 antiferromagnetic spinor BEC, containing up to $1.5\times10^5$
sodium ($^{23}$Na) atoms, in free space at $q/h=42$~Hz. Here $q$ is the quadratic Zeeman energy and $h$ is the Planck
constant. We then apply resonant microwave and imaging pulses to eliminate the atoms in $|F=1,m_F=\pm1\rangle$ states.
This leaves us with a BEC in its superfluid ground state, the longitudinal polar state, where $\rho_0=1$ and $m=0$. After
the pure state is prepared, we quench $q$ to a desired value using magnetic coils and then load the BEC into a cubic
lattice. The cubic lattice is constructed by three orthogonal standing waves producing a lattice with spacing of
$532\,$nm. The lattice beams originate from a single mode laser at $1064\,$nm and are frequency shifted by at least
$20\,$MHz from each other. The lattice depth ($u_L$) is calibrated using Kapitza-Dirac diffraction patterns. Once loaded
we linearly ramp $u_L$ up to the desired value ($u_L^{\rm final}$) in a given ramp time ($t_{\rm ramp}$), which
corresponds to a roughly exponential ramp in the ratio of $U_0/J$. Here $J$ is the hopping energy and $U_0$ is the
spin-independent interaction. The atoms are then abruptly released and ballistically expand, before we measure $\rho_0$
and the superfluid order parameter with a two-stage microwave imaging method~\cite{ZhaoSinglet,ChenQuench}. By holding the
lattice ramp speed ($v_{\rm ramp}=u_L^{\rm final}/t_{\rm ramp}$) constant while varying the $u_L^{\rm final}$ and $t_{\rm
ramp}$, we produce graphs similar to those found in Fig.~2(a) and Fig.~3(a) of the main text. Each experimental data point
represents the average of around 15 shots at a given $u_L^{\rm final}$ and $t_{\rm ramp}$, with error bars representing
one standard deviation of the shots. Each complete set at a given $v_{\rm ramp}$ and $q$ contains 14 or more points at
representative $u_L^{\rm final}$ to allow fitting with either a sigmoid function or a piecewise function given by
$y=y_0+A$ if $x \le x_0$ and $y=y_0+A\cos(f(x-x_0))$ if $x > x_0$. The fitting function used for each set was determined
by the chi-square of the fitted functions and was then used to calculate the critical lattice depth and its associated
error.

The BH Hamiltonian of lattice-trapped $F$=1 spinor gases can be expressed as~\cite{spinorquench},
\begin{eqnarray}
H &=& -J \sum_{\langle i,j \rangle,m_F} \left( a^\dagger_{i,m_F} a^{}_{j,m_F} + a^\dagger_{j,m_F} a^{}_{i,m_F} \right) \nonumber \\
&& + \dfrac{U_0}{2} \sum_i \hat{n}_{i} \left( \hat{n}_{i} -1 \right) + \dfrac{U_2}{2} \sum_i \left( \vec{F}^{2}_{i} -2\hat{n}_{i}\right) - \sum_i \mu_{i} \hat{n}_{i}\nonumber \\
&&+ \frac{1}{2}V_T \sum_i (i-L/2)^2 \hat{n}_{i}+q \sum_i (n_{i,1}+n_{i,-1})~~, \nonumber \\
\label{hamil}
\end{eqnarray}
where $\hat{n_i}=\sum_{m_F} a^\dagger_{i,m_F} a_{i,m_F}$ is the total atom number at site-$i$, $a^\dagger_{m_F}$
($a_{m_F}$) are boson creation (destruction) operators with $m_F$ being $\pm 1$ and 0 for spin-1 atoms, the total spin is
$\vec{F}=\sum_{m_F, m_F'}a^\dagger_{m_F} \vec{F}_{m_F, m_F'} a_{m_F'}$, the standard spin-1 matrices are $\vec{F}_{m_F,
m_F'}$, the spin-dependent interaction is $U_2$, the chemical potential is $\mu$, the total number of lattice sites is
$L$, and $V_T$ is the harmonic trap strength. In a mean-field approximation, the above Hamiltonian can be expressed as a
sum of independent single site Hamiltonians $ H=\sum_{i} H_{i} $ with
\begin{eqnarray}
H_{i} &=& \dfrac{U_0}{2} \hat{n}_{i} \left( \hat{n}_{i} -1 \right) + \dfrac{U_2}{2} \left( \vec{F}^{2}_{i}
-2\hat{n}_{i}\right)\nonumber \\
&&- \mu_i \hat{n}_{i}- zJ\sum_{m_F} \psi_{m_F} \left(a^\dagger_{i,m_F} + a_{i,m_F} \right)\nonumber \\
&&+ q \sum_{m_F} m_F^{2} n_{i,m_F} + zJ\sum_{m_F} |\psi_{m_F}|^2~.\nonumber \\
\label{mfham}
\end{eqnarray}
Here $z$ is the number of nearest neighbors, $\psi_{m_F}=\langle a^{\dagger}_{i,m_F} \rangle=\langle a_{i,m_F} \rangle$ is
the superfluid order parameter, and we absorb the $V_T$ term in the chemical potential. The side peak fraction studied in
Fig.~3 is given by $|\langle \hat{a}_{i,m_F} \rangle|^2$ for any site $i$ and is thus proportional to the superfluid order
parameter.

We compute the static and dynamic properties of the spin-1 BH Hamiltonian using the Gutzwiller
approximation~\cite{Gutzwiller07,Gutzwiller15}, which assumes the many-body wave function in the full lattice can be
written as a product of local states at $L$ individual sites, $|\Psi_{GW} \rangle = \prod^{L}_{i=1}|\phi_i \rangle $ with
$|\phi_i \rangle = \sum_{n_{i, m_F}}A_i(n_{i,1},n_{i,0},n_{i,-1})|n_{i,1},n_{i,0},n_{i,-1} \rangle$. $A_i$ is the
variational parameter at site $i$. Dynamics are derived from numerically solving the time-dependent Schr\"odinger equation
for specific initial states. To perform the Gutzwiller calculations, we write the matrix elements of the Hamiltonian
$H_{i}$ in the occupation number basis $|n_{i,-1},n_{i,0},n_{i,1} \rangle$, and truncate the onsite Hilbert space by
allowing a maximum number of particles per site, $n_{\rm max}=8$, for which dynamics can be computed efficiently and the
truncation effects are negligible.\\

\textbf{Acknowledgments} We thank the National Science Foundation and the Noble Foundation for financial support.

\end{document}